\documentclass[10 pt,aps,prd,superscriptaddress,showkeywords,showpacs,floatfix,nofootinbib]{revtex4-1}
\usepackage{mathtext,bbm,amsmath,amsfonts,amssymb,indentfirst,syntonly,graphicx}
\usepackage[english]{babel}
\usepackage{amsmath,amssymb,amsthm}
\usepackage{latexsym,graphicx,bbm}

\begin{document}
\title{Inflation and primordial power spectra at anisotropic spacetime inspired by Planck's constraints on isotropy of CMB}

\author{Zhe Chang}
\email[]{E-mail: changz@ihep.ac.cn}
\author{Sai Wang}
\email[]{E-mail: wangsai@ihep.ac.cn}

\affiliation{Institute of High Energy Physics, Chinese Academy of Sciences, 100049 Beijing, China}
\affiliation{Theoretical Physics Center for Science Facilities, Chinese Academy of Sciences, 100049 Beijing, China}

\begin{abstract}
Recently, the Planck 2013 results showed possible evidence for a dipolar power modulation of the CMB temperature fluctuations at low--\(\ell\) multipoles. This anomaly might imply certain deviations from statistical isotropy. To incorporate the Planck's data into standard cosmological model, we propose an inflation model of the very early universe in an anisotropic spacetime. A generalized Friedmann-Robertson-Walker (FRW) metric is presented in the Randers-Finsler spacetime. We obtain the primordial power spectrum of the scalar perturbation with direction dependence, such as the dipolar modulation. This is consistent with the dipolar power modulation of the CMB anisotropy signaled by the Planck observation.
\end{abstract}
%\pacs{}
%\keywords{Finsler geometry, Inflation, Primordial power spectra}
\maketitle

\section{Introduction}\label{sec:introduction}

The cosmological principle assumes that the universe is statistically isotropic and homogeneous at large scales \cite{Book by Dodelson}.
Recently, it has been rigorously tested by the observations on the CMB anisotropy from the Planck satellite \cite{Planck 2013 resultsI}
and the Wilkinson Microwave Anisotropy Probe (WMAP) \cite{WMAP7,WMAP9}.
It was showed that the statistical isotropy is well consistent with these datasets.
However, there are several anomalies at low--\(\ell\) multipoles of the CMB temperature fluctuations \cite{WMAP9,WMAP717,Planck2013resultsXXIII}.
They include the quadrupole-octopole alignment \cite{quadrupole-octopole alignment,mode alignment by Copi},
the hemispherical asymmetry \cite{hemispherical asymmetry01,hemispherical asymmetry02,hemispherical asymmetry03},
the dipolar power modulation \cite{dipolar power modulation01,dipolar power modulation02,dipolar power modulation03},
the parity asymmetry \cite{parity asymmetry01,parity asymmetry02,parity asymmetry03,parity asymmetry04,parity asymmetry05,Dipole01,parity asymmetry06},
the missing power of quadrupole \cite{missing power001,missing power002},
the cold spot \cite{cold spot001,cold spot002,cold spot003,cold spot004,cold spot005,cold spot006,cold spot007},
and so on.
These anomalies give rise to extensively theoretical debates recently.
It is believed that they might be stemmed from the non-standard spinor \cite{nonstandard spinor},
the curvaton field \cite{curvaton field01,curvaton field02},
the primordial vector fields \cite{DurrerKY1998,KimN2009,vector field02,vector field03,Cosmological Magnetic Fields01,
Cosmological Magnetic Fields02,Cosmological Magnetic Fields03,Cosmological Magnetic Fields04,Cosmological Magnetic Fields05,
primordial preferred direction001,The statistically anisotropic curvature perturbation 1302}, and so on.

Originally, the hemispherical asymmetry was tested by studying the angular power spectrum locally at different positions
on the celestial sphere \cite{hemispherical asymmetry01,hemispherical asymmetry02}.
It was found that the power spectrum for a hemisphere was larger than that for the opposite hemisphere at low--\(\ell\) multipoles.
This power asymmetry might take the form of dipole modulation over the sky \cite{WMAP717}, namely
\begin{equation}
\label{dipolar power modulation}
\Delta T(\hat{\textbf{p}})=\left(1+A\hat{\textbf{p}}\cdot \hat{\textbf{n}}\right)\Delta T_{iso}(\hat{\textbf{p}})\ ,
\end{equation}
where \(\Delta T_{iso}\) denotes the unmodulated isotropic quantity,
\(\hat{\textbf{n}}\) is the dipole direction and \(\hat{\textbf{p}}\) is a unit direction in the sky.
Based on the WMAP data, the dipole amplitude was found to be \(A=0.07\pm0.02\) for \(\ell<64\).
The dipole direction was \((l,b)=(224^\circ \pm24^\circ,-22^\circ \pm24^\circ)\).
Recently, the Planck 2013 results \cite{Planck2013resultsXXIII} confirmed the above asymmetry of the CMB at the level of around \(3\sigma\).
For the NILC map, for example, the dipole amplitude referred to \(A=0.07\pm0.01\)
and the dipole direction was \((l,b)=(218.9^\circ \pm15.4^\circ,-21.4^\circ \pm15.1^\circ)\).
Similar results were obtained for the other three maps.
These are consistent with the WMAP--based results.

To incorporate the dipolar power modulation of the CMB anisotropy into the standard cosmological model,
we could investigate the inflation with anisotropy.
Phenomenologically, the primordial power spectrum would be of direction dependence in the anisotropic inflation.
It could be written as \cite{curvaton field02}
\begin{equation}
\label{primordial power spectrum with direction dependence}
\mathcal{P}_{\mathcal{R}}^{1/2}(\textbf{k},\textbf{r})=\left(1+A(k,r)\hat{\textbf{r}}\cdot\hat{\textbf{n}}\right)\mathcal{P}_{\mathcal{R}}^{1/2}(k)\ ,
\end{equation}
where \(\mathcal{P}_{\mathcal{R}}(k)\) denotes the isotropic power spectrum.
Here \(A(k,r)\) is a positive function of the wavenumber \(k\) and the distance \(r\).
We could relate \(r\sim k^{-1}\), since the CMB at the scale \(r\) is most sensitive to the perturbation with \(k\sim r^{-1}\).
The parameter \(A\) would be of the level \(0.07\) at the scales of \(k^{-1}\sim1\rm{Gpc}\) \cite{Planck2013resultsXXIII},
while it would be less than \(0.015\) at the scales of \(k^{-1}\sim1\rm{Mpc}\) \cite{Constraints on cosmic hemispherical power anomalies from quasars}.

In this paper, we would study such an inflation model in an anisotropic spacetime.
The natural framework of anisotropic spacetime is the so-called Finsler geometry \cite{Finsler}.
Finsler geometry gets rid of the quadratic constraint on the metric \cite{Book by Rund,Book by Bao,Book by Shen}.
The Finsler spacetime admits certain privileged axes and permits less symmetries than the Riemann one does \cite{Finsler isometry by Wang,Finsler isometry by Rutz,Finsler isometry LiCM}.
Thus, the Finsler geometry is a reasonable candidate to reveal the deviations from isotropy of the spacetime.
For instance, the Finsler spacetime could account for the Lorentz violation as well as the CPT violation \cite{A special-relativistic theory of the locally anisotropic space-time I,A special-relativistic theory of the locally anisotropic space-time II,A special-relativistic theory of the locally anisotropic space-time Appendix,recoiling D-branes02,DSR in Finsler,VSR in Finsler,Kostelecky_Finsler,ChangWangepjc 2012,ChangWangepjc 2013,ChangLiWangepjc 2012}.
The cosmic acceleration could be explained by the anisotropic Friedmann equation in Finsler cosmology \cite{GVSR in Finsler cosmology}.
The large-scale bulk flow \cite{Kashlinsky0809,Kashlinsky0910} could be revealed by a Finslerian Zermelo navigation model \cite{Chang et al 2013a}.
The Randers--Finsler structure could account for the privileged direction \cite{AntoniouPerivolaropous2010,Direction Dependence of the Deceleration Parameter,Constraints on anisotropic cosmic expansion from supernovae,Jackson2012} of the maximum accelerating expansion in the Hubble diagram \cite{Chang et al 2013b}.
A spatially anisotropic Finslerian model could account for the mass discrepancy problem of the Bullet cluster \cite{Lietal2012}, and so on.

The statistical anisotropy of the CMB may be related with the Randers--Finsler spacetime.
The Randers structure \cite{Randers space} involves an extra 1-form which is related to a vector field.
This vector field may influence the very early evolution of the universe if it existed in the universe.
In this paper, we propose a generalized FRW metric in the Randers spacetime.
The Randers structure comprises the FRW part and the weak vector field.
The vector field singles out a privileged axis in the very early universe.
We study Einstein's gravitational field equations via the osculating Riemannian approach.
The time-time component of Einstein's field equations is resolved to obtain an inflationary phase of the very early universe.
The anisotropic modifications are studied for the inflationary universe.
By studying the equation of motion for the inflaton field,
we obtain the primordial power spectrum of the scalar perturbation with statistical anisotropy.
The predicted statistical anisotropy may account for the Planck observed dipole modulation of the CMB temperature fluctuations.

The rest of the paper is arranged as follows.
In section \ref{sec:Finsler}, we first present a brief introduction to Finsler geometry.
In section \ref{sec:Randersfieldequations}, the osculating Riemannian approach is introduced to study a generalized FRW metric in the Randers spacetime.
In section \ref{sec:Inflation}, we study the time-time component of Einstein's gravitational field equations
and resolve it to obtain an inflationary solution.
Meanwhile, the anisotropic modifications are acquired for the inflationary phase of the universe.
The primordial power spectrum with statistical anisotropy is got through the equation of motion for the inflaton field in section \ref{sec:Primordialpowerspectra}.
Conclusions and discussions are listed in section \ref{sec:Conclusions}.

\section{The anisotropic spacetime}\label{sec:Finsler}

In this paper, we suggest the Randers spacetime as a suitable background of inflation at the very early stage of the universe.
The Randers spacetime is a class of Finsler spacetimes.
Thus, we first present a brief introduction to the Finsler geometry in this section.
For more detailed discussions on Finsler geometry, see the references, for instance \cite{Book by Rund,Book by Bao,Book by Shen}.

Different from Riemann geometry, Finsler geometry is defined on the tangent bundle.
Finsler space stems from the arc-length integrals of the form
\begin{equation}
\label{Finsler origin}
s=\int_a^b F(x,y)ds\ ,
\end{equation}
where \(x\) and \(y\equiv dx/ds\) denote the location and the velocity, respectively.
The proper length is given by \(s\).
The integrand \(F(x,y)\) is called the Finsler structure.
It is a smooth and positive function on the tangent bundle.
It is positively homogeneous of degree one, i.e.,
\(F(x,\lambda y)=\lambda F(x,y)\) for \(\lambda>0\).
The Finsler metric is defined by the Hessian,
\begin{equation}
\label{Finsler metric}
g_{\mu\nu}=\frac{\partial}{\partial y^{\mu}}\frac{\partial}{\partial y^{\nu}}\left(\frac{1}{2}F^{2}\right)\ .
\end{equation}
Together with its inverse \(g^{\mu\nu}\), it is used to lower and raise the spatial indices of tensors.
For the Finsler spacetime, the spatial indices run from 1 to 3 and the temporal index runs 0 in this paper.

The Finsler metric is dependent on the positions \(x\) as well as the directions given by the fibre \(y\).
This is different from the Riemann metric which depends on \(x\) only.
Thus, there is other significant object, i.e., the Cartan (torsion) tensor in Finsler geometry.
The Cartan tensor is defined as \cite{Book by Bao}
\begin{equation}
\label{Cartan tensor}
C_{\mu\nu\sigma}=\frac{1}{2}\frac{\partial g_{\mu\nu}}{\partial y^{\sigma}}\ .
\end{equation}
It completely characterizes the deviation of the Finsler space from the Riemann space.
The Finsler space becomes Riemannian if and only if \(C_{\mu\nu\sigma}=0\).
This result reveals that Finsler geometry is a natural generalization of Riemann geometry.

The Randers space \cite{Randers space} is a kind of Finsler spaces.
It could be viewed as a Riemann space influenced by a primordial vector field (for instance, the electromagnetic field).
The Randers structure is given by
\begin{equation}
\label{Randers structure}
F(x,y)=\alpha(x,y)+\beta(x,y)\ ,
\end{equation}
where \(\alpha(x,y)=\sqrt{\tilde{a}_{\mu\nu}(x)y^{\mu}y^{\nu}}\) denotes a Riemann structure and \(\beta(x,y)=\tilde{b}_{\mu}(x)y^{\mu}\) is a 1-form.
However, the vector field appears as a part of the space structure in the Randers space.
Furthermore, it induces the anisotropic properties of the Randers space.
The tildes here denote raising (or lowering) the indices with the Riemann metric \(\tilde{a}_{\mu\nu}\) (or \(\tilde{a}^{\mu\nu}\)).
To reveal the anisotropy of the Randers space,
one notices that the Randers structure is not absolutely homogeneous of degree one.
Otherwise, the Randers structure reduces back to the Riemann one.
The reason is that \(\alpha\) would keep invariant while \(\beta\) changes its sign under the rescale \(y\longrightarrow -y\).
In addition, the anisotropy of the Finsler spaces could also be revealed by the number of solutions of the Killing equations \cite{Finsler isometry by Wang,Finsler isometry by Rutz,Finsler isometry LiCM}.

In the Randers space, the Randers metric is defined as \cite{Book by Bao}
\begin{equation}
\label{Randers metric}
g_{\mu\nu}=\frac{F}{\alpha}\left(\tilde{a}_{\mu\nu}-\tilde{\ell}_{\mu}\tilde{\ell}_{\nu}\right)+\ell_{\mu}\ell_{\nu}\ ,
\end{equation}
where \(\ell_{\mu}=\tilde{\ell}_{\mu}+\tilde{b}_{\mu}\) and \(\tilde{\ell}_{\mu}={\tilde{a}_{\mu\nu}y^{\nu}}/{\alpha}\).
Its inverse is given by
\begin{equation}
\label{Randers metric inverse}
g^{\mu\nu}=\frac{\alpha}{F}\tilde{a}^{\mu\nu}+\frac{\alpha^2}{F^2}\frac{\beta+\alpha\|\tilde{b}\|^2}{F}\tilde{\ell}^{\mu}\tilde{\ell}^{\nu}
-\frac{\alpha^2}{F^2}\left(\tilde{\ell}^{\mu}\tilde{b}^{\nu}+\tilde{\ell}^{\nu}\tilde{b}^{\mu}\right)\ ,
\end{equation}
where \(\|b\|^2=\tilde{b}^{\mu}\tilde{b}_{\mu}\).
The Cartan tensor of Randers space is given as \cite{Book by Bao}
\begin{equation}
\label{Cartan tensor in Randers}
C_{\mu\nu\sigma}=\frac{1}{2\alpha}\mathcal{S}_{(\mu\nu\sigma)}\left(\tilde{a}_{\mu\nu}-\tilde{\ell}_{\mu}\tilde{\ell}_{\nu}\right)
\left(\tilde{b}_{\sigma}-\frac{\beta}{\alpha}\tilde{\ell}_{\sigma}\right)\ ,
\end{equation}
where \(\mathcal{S}_{(\mu\nu\sigma)}\) refers the summation over the cyclic permutation of indices.
We note that each term in \(C_{\mu\nu\sigma}\) is proportional to the components of \(\tilde{b}\).
Thus, the Cartan tensor could be a first order object in the case of \(\|\tilde{b}\|\ll 1\).

\section{The osculating Riemannian approach}\label{sec:Randersfieldequations}

In this section, we consider a generalized FRW metric with weak anisotropy in the Randers spacetime.
Following Stavrinos {\it et al.}'s approach \cite{FRW model with weak anisotropy by Stavrinos}, we will employ the osculating Riemannian metric to approximate the Randers metric.

The related Randers structure is given by the spatially flat FRW metric \(\tilde{a}_{\mu\nu}=diag(1,-a^2(t),-a^2(t),-a^2(t))\)
and an extra 1-form \(\tilde{b}_{\mu}dx^{\mu}\) (or a vector field \(\tilde{b}^{\mu}{\partial}/{\partial x^{\mu}}\)).
The temporal coordinate denotes the proper time measured by a comoving observer while the spatial coordinates are comoving.
The velocity \(y^{\mu}=(1,0,0,0)\) denotes the tangent 4-velocity of the comoving observer along a worldline.
%Here \(\tau\) denotes the proper time.
The vector field is very weak, i.e., \(\|\tilde{b}\|\ll 1\) in the Randers spacetime.
It would be expected to align with the tangent 4-velocity \cite{Book by Misner}.
Thus, we consider the timelike vector field that has only the temporal component non-vanishing,
i.e., \(\tilde{b}_{\mu}=(B(z),0,0,0)\).
In addition, the component \(B(z)\) is set to be dependent on the third spatial coordinate \(z\) only.
This proposition would induces the anisotropy of the primordial power spectrum along the \(z\)-axis.
Therefore, we will study the generalized FRW metric (i.e., the Randers structure) as
\begin{equation}
\label{GFRW structure in Randers}
d\tau=\sqrt{dt^2-a^2(t)\left(dx^2+dy^2+dz^2\right)}+B(z)dt\ ,
\end{equation}
where the spatially flat FRW part is given by the first term on the right hand side.
The anisotropy of the obtained Randers spacetime is characterized by the second term on the right hand side of the above equation.
For the weak vector field, the Randers metric could be viewed as a slight modification to the FRW metric.

Usually, it is difficult to discuss the gravitational issues in the completely Finsler geometric framework.
However, the Finsler metric could be approximately related to the Riemann metric in certain cases \cite{Book by Rund,Book by Asanov}.
Actually, it is convenient to study the gravity with the osculating Riemannian method \cite{Book by Rund}.
The Finsler structure corresponds to the osculating Riemannian metric as
\begin{equation}
\label{Randers metric for B(z)}
g_{\mu\nu}(x)\equiv g_{\mu\nu}(x,y(x))\ ,
\end{equation}
where the velocity \(y(x)\) is viewed as a function of the position \(x\).
%Then the osculating Riemannian structure becomes \(ds^2=\left(1+B(z)\right)^2 dt^2-a^2(t)\left(1+B(z)\right)\left(dx^2+dy^2+dz^2\right)\)
%for the above generalized FRW metric (\ref{GFRW structure in Randers}) in the Randers spacetime.
Correspondingly, the coefficients of Christoffel symbol are given by \cite{FRW model with weak anisotropy by Stavrinos}
\begin{eqnarray}
\label{Christoffel symbols}
\gamma^{\mu}_{\nu\sigma}(x)=&&\gamma^{\mu}_{\nu\sigma}(x,y(x))+C^{\mu}_{\nu\lambda}(x,y(x))\frac{\partial y^{\lambda}}{\partial x^{\sigma}}(x)\nonumber\\
&+&C^{\mu}_{\lambda\sigma}(x,y(x))\frac{\partial y^{\lambda}}{\partial x^{\nu}}(x)\nonumber\\
&-&g^{\mu\rho}(x,y(x))C_{\kappa\nu\sigma}(x,y(x))\frac{\partial y^{\kappa}}{\partial x^{\rho}}(x)\ ,
\end{eqnarray}
where \(\gamma^{\mu}_{\nu\sigma}(x,y(x))=\frac{g^{\mu\kappa}}{2}\left(\frac{\partial g_{\kappa\sigma}}{\partial x^{\nu}}-\frac{\partial g_{\nu\sigma}}{\partial x^{\kappa}}+\frac{\partial g_{\nu\kappa}}{\partial x^{\sigma}}\right)\).
From the equation (\ref{Cartan tensor in Randers}), we may note that the Cartan tensor
is proportional to the components of the vector \(\tilde{b}\) which is small.
In addition, we have chosen the comoving coordinates in the Randers spacetime.
Thus, the terms dependent on the Cartan tensor could be dropped in the equation (\ref{Christoffel symbols}).
In the Randers spacetime, therefore, we approximately obtain the Christoffel symbols as
\begin{equation}
\label{Christoffel symbols approximately}
\Gamma^{\mu}_{\nu\sigma}(x)\approx \gamma^{\mu}_{\nu\sigma}(x,y(x))\ .
\end{equation}
Here \(\Gamma^{\mu}_{\nu\sigma}(x)\) denote the coefficients of the osculating Christoffel connection.

Corresponding to the osculating Christoffel connection, the components of the curvature tensor are given by
\begin{equation}
\label{curvature}
R^{~\mu}_{\nu~\rho\sigma}=\Gamma^{\mu}_{\nu\sigma,\rho}-\Gamma^{\mu}_{\nu\rho,\sigma}+\Gamma^{\kappa}_{\nu\sigma}\Gamma^{\mu}_{\kappa\rho}
-\Gamma^{\kappa}_{\nu\rho}\Gamma^{\mu}_{\kappa\sigma}\ .
\end{equation}
The Ricci tensor and the scalar curvature are, respectively, given as
\begin{eqnarray}
\label{Ricci tensor}
Ric_{\mu\nu}&=&R^{~\kappa}_{\mu~\kappa\nu}\ ,\\
\label{Scalar curvature}
S&=&g^{\mu\nu}(x)Ric_{\mu\nu}\ .
\end{eqnarray}
The osculating Riemannian metric evolves following the conventional Einstein's gravitational field equations,
\begin{equation}
\label{Einstein's field equations}
Ric_{\mu\nu}-\frac{1}{2}g_{\mu\nu}S=8\pi G T_{\mu\nu}\ ,
\end{equation}
where \(T_{\mu\nu}=\partial_\mu \phi\,\partial_\nu \phi \,-\,
g_{\mu\nu}\left( \frac{1}{2}\,g^{\alpha\beta}\,\partial_\alpha
\phi\,\partial_\beta \phi \,-\, V(\phi)\right)\) denotes the energy-momentum tensor of the cosmic fluid.
However, the anisotropic properties of the Finsler spacetime have been approximately comprised into the osculating Riemannian metric.
Thus, the obtained Einstein's field equations are weakly anisotropic in the Randers spacetime considered in this paper.

\section{The inflationary phase with weak anisotropy}\label{sec:Inflation}

In this section, we study Einstein's gravitational field equations in the Randers spacetime.
The inflationary solution is obtained for the very early universe at the zero-order assumption.
Further, we take account the slight modifications to the inflationary phase such that the very early spacetime acquires the weak anisotropy.

First, we calculate the time-time component of the Einstein tensor to study Einstein's field equations in the Randers spacetime.
The Einstein tensor is defined as \(G_{\mu\nu}=Ric_{\mu\nu}-\frac{1}{2}g_{\mu\nu}S\).
We obtain its time-time component as
\begin{equation}
\label{Einstein tensor}
G_{00}=3\left(\frac{\dot{a}}{a}\right)^2+\left(\frac{1}{a}\right)^2\frac{3B'^2-4B''(1+B)}{4(1+B)}\ ,
\end{equation}
where the dots denote the derivative of \(a\) with respective to \(t\) while the primes denote the derivative of \(B\) with respective to \(z\).
For an inflationary phase \cite{Inflation by Starobinsky,Inflation by Guth,Inflation by Linde,Inflation by Steinhardt,Inflation by Linde0}, the very early universe undergoes a process of exponential expansion,
i.e., \(a\sim e^{Ht}\) where \(H\) is the Hubble horizon.
Thus, the second term on the right hand side of the equation (\ref{Einstein tensor}) would decrease exponentially.
It could be discarded at the zero-order approximation.
In this way, Einstein's field equation would reduce back to the conventional one.
To account for the spacetime anisotropy, we could consider the modifications from the second term on the right hand side of (\ref{Einstein tensor}).

At the zero-order approximation, the time-time component of the Einstein tensor becomes
\begin{equation}
\label{Einstein tensor first order}
G_{00}^{(0)}= 3\left(\frac{\dot{a}}{a}\right)^2\ .
\end{equation}
On the other hand, the energy-momentum tensor of the cosmic fluid is characterized by the one of the perfect fluid.
At the zero order, its time-time component is given by \cite{Book by Dodelson}
\begin{equation}
\label{Energy density first order}
T_{00}^{(0)}=\rho^{(0)}=\frac{1}{2}\left(\frac{d\phi^{(0)}}{dt}\right)^2+V^{(0)}(\phi^{(0)})\ ,
\end{equation}
where \(\phi^{(0)}(t)\) is the zero-order part of the mostly homogeneous inflaton field \(\phi(t,\vec{x})\).
Consider the slow-roll condition that the inflaton field slowly rolls down its potential,
i.e., \(\left(\frac{d\phi^{(0)}}{dt}\right)^2\ll V^{(0)}(\phi^{(0)})\).
Thus, the energy density \(T_{00}^{(0)}\) of the inflaton field almost becomes a constant.
The time-time component of Einstein's equations becomes \cite{Book by Dodelson}
\begin{equation}
\label{Einstein field equations for Slow-rolling}
\left(\frac{\dot{a}}{a}\right)^2=\frac{8\pi G}{3}V^{(0)}(\phi^{(0)})=const.\ ,
\end{equation}
which determines the evolution of scale factor \(a\).
This equation has an exponential solution \(a(t)\sim e^{Ht}\) where \(H\) denotes the Hubble horizon which is almost constant.
The exponential expansion of the very early universe corresponds to the inflationary phase of the universe .
Note that the above analysis is similar to that in the standard inflationary model, see \cite{Book by Dodelson} for details.

To account for the anisotropic effects,
we substitute the exponential solution \(a(t)\sim e^{Ht}\) back into Einstein's field equation (\ref{Einstein's field equations})
and obtain the anisotropic modification \(B(z)\) of the very early universe.
In this case, the time-time component of the Einstein tensor is given by the equation (\ref{Einstein tensor}),
and the energy density of the inflaton field is given by \(T_{00}=\frac{1}{2}\left(\frac{d\phi^{(0)}}{dt}\right)^2+(1+B)^2 V(\phi^{(0)})\).
Here the potential of the inflaton field is set to be \(V(\phi^{(0)})=V^{(0)}(\phi^{(0)})/\left(1+B\right)^2\),
where the correction \((1+B)^2\) in \(T_{00}\) comes from the osculating Riemannian metric \(g_{00}\).
Then the time-time component of Einstein's field equation (\ref{Einstein's field equations}) becomes
\begin{equation}
\label{modifications to Einstein's field equation}
3B'^2-4B''(1+B)=0\ ,
\end{equation}
where we have used the equations (\ref{Einstein tensor}), (\ref{Einstein tensor first order}), (\ref{Energy density first order})
and (\ref{Einstein field equations for Slow-rolling}) in calculations.
We see that the remained term in the equation (\ref{modifications to Einstein's field equation})
determines the anisotropic modifications to the inflationary phase of the universe.
The equation (\ref{modifications to Einstein's field equation}) has a solution as
\begin{equation}
\label{B(z) solution}
B(z)=\frac{1}{256}\left(c_{1}c_{2}\right)^{4}\left(1+\frac{z}{c_{2}}\right)^{4}-1\ ,
\end{equation}
where \(c_1\) and \(c_2\) are the integral constants.
In addition, \(c_{2}>0\) denotes a given distance scale which could be constrained by cosmological observations.
We could set \(|c_{1}c_{2}|\approx 4\) and \(|z|\ll c_{2}\) because of the condition \(|B(z)|\ll 1\).
Note that \(B(z)\) would be a monotonically increasing function of \(z\).
To the second-order approximation, the above solution \(B(z)\) could be expanded as \(B(z)\simeq f_{0}+f_{1}z+\frac{3f_{1}^{2}}{8(1+f_{0})^{2}}z^{2}+o(z^{2})\) where \(f_{0}=|c_{1}c_{2}|-4\) and \(f_{1}=4(1+f_{0})/c_{2}\).
In this way, we obtain the inflationary phase of the generalized FRW metric (\ref{GFRW structure in Randers})
with weak anisotropy in the Randers spacetime.

\section{The primordial power spectrum with the direction dependence}\label{sec:Primordialpowerspectra}

In this section, we study the primordial power spectrum \cite{Fluctuation01,Fluctuation02,Fluctuation03,Fluctuation04,Fluctuation05,Fluctuation06,Fluctuation07}
with weak anisotropy in the Randers spacetime.
First, the action of the inflaton field is given by
\begin{equation}
\label{Action of Inflaton}
S=\int d^4x\sqrt{-g}\mathcal{L}=\int d^4x\sqrt{-g}\left(\frac{1}{2}g^{\mu\nu}\partial_{\mu}\phi\partial_{\nu}\phi-V(\phi)\right)\
\end{equation}
where \(\sqrt{-g}=det(g_{\mu\nu})\) for the osculating Riemannian metric of the Randers structure (\ref{GFRW structure in Randers}).
From the Euler-Lagrange equation
\(\sqrt{-g}\frac{\partial \mathcal{L}}{\partial \phi}-\partial_{\mu}
\left(\sqrt{-g}\frac{\partial \mathcal{L}}{\partial (\partial_{\mu}\phi)}\right)=0\),
we obtain the equation of motion of the inflaton field as
\begin{equation}
\label{Equation of motion for inflaton}
\ddot{\phi}+3H\dot{\phi}-\frac{1+B}{a^2}\nabla^2 \phi+\frac{dV^{(0)}(\phi)}{d\phi}-\frac{3}{2}\frac{B'}{a^2}\phi'=0\ .
\end{equation}
We have decomposed the inflaton field into two parts \(\phi(t,\textbf{x})=\phi^{(0)}(t)+\delta\phi(t,\textbf{x})\),
where \(\delta\phi(t,\textbf{x})\) denotes a first-order perturbation.

For the zero-order part \(\phi^{(0)}(t)\), the equation of motion becomes
\begin{equation}
\label{equation of motion zero order}
\ddot{\phi}^{(0)}+3H\dot{\phi}^{(0)}+\frac{dV^{(0)}}{d\phi}|_{\phi=\phi^{(0)}}=0\ .
\end{equation}
For the anisotropic and inhomogeneous perturbation, the equation of motion for the fluctuations becomes
\begin{equation}
\label{equation of motion first order}
\ddot{\delta\phi}+3H\dot{\delta\phi}+\frac{d^2 V^{(0)}}{d\phi^{2}}|_{\phi=\phi^{(0)}}\delta\phi-\frac{1+B}{a^2}\nabla^2\delta\phi-\frac{3}{2}\frac{B'}{a^2}\delta\phi'=0\ ,
\end{equation}
where we have used the equation (\ref{equation of motion zero order}) to eliminate \(\phi^{(0)}\).
Typically, the \(\frac{d^2 V^{(0)}}{d\phi^{2}}\) term is small, which is proportional to the slow-roll variable \(\eta\simeq \frac{d^2 V^{(0)}}{d\phi^{2}}/(3H^{2})\).
It could be comparable to the magnitude of the anisotropic term.
However, this term just affects the spectral index \(n_{\delta\phi}\)
\footnote{In this case, the equation of motion (\ref{Fluctuations conformal factor}) becomes
\begin{equation}
\frac{d^2 \delta\sigma_{\textbf{k}}}{d\tau^2}+\left(k_{eff}^{2}-\frac{2-3\eta}{\tau^2}\right)\delta\sigma_{\textbf{k}}=0\ .\nonumber
\end{equation}
Thus, only the spectral index would be affected based on this equation.
} \cite{Lecture by Riotto}.
The isotropic power spectrum \(\mathcal{P}_{\mathcal{\delta\phi}}^{iso}(k)\) would be affected.
Thus, we could neglect it in the following discussions if we are only interested in the anisotropic effect.

The fluctuation of inflaton field could be expanded into the Fourier modes as
\begin{equation}
\label{Fourier modes}
\delta{\phi(t,\textbf{x})}=\int\frac{d^3 \textbf{k}}{(2\pi)^{3/2}}e^{i\textbf{k}\cdot\textbf{x}}\delta\phi_{\textbf{k}}(t)\ .
\end{equation}
In this way, the fluctuations would evolve along the equation of motion as
\begin{equation}
\label{Fluctuations motion Fourier modes}
\ddot{\delta\phi_{\textbf{k}}}+3H\dot{\delta\phi_{\textbf{k}}}+\frac{k_{eff}^{2}}{a^2}\delta\phi_{\textbf{k}}=0\ ,
\end{equation}
where the effective wavenumber is given as
\begin{equation}
\label{keff}
k_{eff}^{2}=k^{2}\left(1+B-i\frac{3B'}{2k}(\hat{\textbf{k}}\cdot\hat{\textbf{n}})\right)\ ,
\end{equation}
and and \(\hat{\textbf{k}}\cdot\hat{\textbf{n}}\) denotes the cosine of angle between the wavevector \(\textbf{k}\) and the third spatial direction \(\hat{\textbf{n}}\).

With the redefinition of the inflaton \(\delta\phi_{\textbf{k}}=\delta\sigma_{\textbf{k}}/a\),
we could work in the conformal time \(d\tau=dt/a\) where the coefficient \((1+B)^2\) is discarded
since it does not affect the following discussions.
Thus, the equation of motion of the fluctuations becomes
\begin{equation}
\label{Fluctuations motion Fourier modes conformal}
\frac{d^2 \delta\sigma_{\textbf{k}}}{d\tau^2}+\left(k_{eff}^{2}-\frac{1}{a}\frac{d^2a}{d\tau^2}\right)\delta\sigma_{\textbf{k}}=0\ .
\end{equation}
For the inflationary phase of the universe, the scale factor expands nearly exponentially, i.e., \(a(t)\sim e^{Ht}\).
Correspondingly, the conformal scale factor reads
\begin{equation}
\label{conformal factor}
a(\tau)\simeq-\frac{1}{H\tau}~~(\tau<0)\ .
\end{equation}
Thus, we find that the equation (\ref{Fluctuations motion Fourier modes conformal}) becomes as
\begin{equation}
\label{Fluctuations conformal factor}
\frac{d^2 \delta\sigma_{\textbf{k}}}{d\tau^2}+\left(k_{eff}^{2}-\frac{2}{\tau^2}\right)\delta\sigma_{\textbf{k}}=0\ .
\end{equation}
This equation has an exact solution \cite{Book by Dodelson}
\begin{equation}
\label{Fluctuations exact solution sigma}
\delta\sigma_{\textbf{k}}=\frac{e^{ik_{eff}\tau}}{\sqrt{2k_{eff}}}\left(1+\frac{i}{k_{eff}\tau}\right)\ .
\end{equation}
By considering the redefinition of \(\delta\phi_{\textbf{k}}\), we obtain
\begin{eqnarray}
\label{Fluctuations exact solution phi}
|\delta\phi_{\textbf{k}}|^2&=&\frac{1}{a^2}\frac{1}{2|k_{eff}|}\left(1+\frac{1}{(|k_{eff}|\tau)^2}\right)\nonumber\\
&=&\frac{H^2\tau^2}{2|k_{eff}|}\left(1+\frac{1}{(|k_{eff}|\tau)^2}\right)\ .
\end{eqnarray}

The primordial power spectrum of \(\delta\phi_{\textbf{k}}\), which is denoted by \(\mathcal{P}_{\delta\phi_{\textbf{k}}}\), is defined by
\begin{equation}
\label{primordial power spectrum of phi}
\langle \delta\phi_{\textbf{k}}^{\ast}\delta\phi_{\textbf{k}'} \rangle=
\delta^{(3)}(\textbf{k}-\textbf{k}')\frac{2\pi^2}{k^{3}}\mathcal{P}_{\delta\phi_{\textbf{k}}}\ .
\end{equation}
Thus, one has the primordial power spectrum of \(\delta\phi_{\textbf{k}}\) as
\begin{eqnarray}
\label{Primordial spectrum phi}
\mathcal{P}_{\delta\phi_{\textbf{k}}}&=&\frac{k^3}{2\pi^2}|\delta\phi_{\textbf{k}}|^2\nonumber\\
&=&\left(\frac{H}{2\pi}\right)^{2}\left(\frac{k}{|k_{eff}|}\right)^{3}+\left(\frac{H}{2\pi}\right)^{2}\frac{k}{|k_{eff}|}\left(k\tau\right)^{2}\ .
\end{eqnarray}
For the super-horizon perturbations, the wavelength \(k\) is much larger than the horizon, i.e., \(k\tau\ll 1\).
Thus, the primordial power spectrum of \(\delta\phi_{\textbf{k}}\) approximately becomes
\begin{equation}
\label{primordial power spectrum of phi approximate}
\mathcal{P}_{\delta\phi_{\textbf{k}}}=\left(\frac{H}{2\pi}\right)^{2}\left(\frac{k}{|k_{eff}|}\right)^{3}\ .
\end{equation}
Note that \(|k_{eff}|^2=k^{2}\left((1+B)^{2}+(3B'/2k)^{2}(\hat{\textbf{k}}\cdot\hat{\textbf{n}})^{2}\right)\) contains the anisotropic properties.

On the super-horizon scales, the primordial power spectrum of the comoving curvature perturbation \(\mathcal{R}\)
is given by \cite{Lecture by Riotto}
\begin{equation}
\label{Primordial power spectrum of R}
\mathcal{P}_{\mathcal{R}}(\textbf{k},z)=\left(\frac{H}{\dot{\phi}}\right)^{2}\mathcal{P}_{\delta\phi_{\textbf{k}}}=
\frac{1}{2m_{pl}^{2}\epsilon}\left(\frac{H}{2\pi}\right)^{2}\left(\frac{k}{|k_{eff}|}\right)^{3}\ ,
\end{equation}
where \(m_{pl}\) is the Planck mass.
The above equation could be parameterized as
\begin{equation}
\label{approximate parametrization}
\mathcal{P}_{\mathcal{R}}(\textbf{k},\textbf{r})\equiv \mathcal{P}_{\mathcal{R}}^{iso}(k)
\left(1-3B+6B^2-\frac{3}{2}\left(\frac{3B'}{2k}\right)^{2}(\hat{\textbf{k}}\cdot\hat{\textbf{n}})^{2}\right)\ ,
\end{equation}
where \(\mathcal{P}_{\mathcal{R}}^{iso}(k)\equiv A_{\mathcal{R}}\left({k}/{H}\right)^{n_{\mathcal{R}}-1}\),
\(A_{\mathcal{R}}\) denotes the normalized amplitude
and \(n_{\mathcal{R}}\) denotes the spectral index of the comoving curvature perturbation.
We remain only the terms that are not higher than \(B^{2}\) orders.
%In addition, \(B\) could be given by \(B(z\rightarrow 2\pi k^{-1}(\hat{k}\cdot\hat{n}))\) because of
%\(z\sim\lambda_{z}=\lambda(\hat{k}\cdot\hat{n})=2\pi k^{-1}(\hat{k}\cdot\hat{n})\).
The terms containing \(B\) induce the anisotropic behaviors of
the primordial power spectrum of the comoving curvature perturbation.
The reason is that  \(B=B(\textbf{r}\cdot\hat{\textbf{n}})\) depends on the third spatial coordinate \(z=\textbf{r}\cdot\hat{\textbf{n}}\).
This term would give arise to the (parity violating) statistical anisotropy of the universe.
Here \(\hat{\textbf{n}}\) denotes the third spatial direction and \(\textbf{r}\) is the spatial location.
It was noteworthy that the level of the statistically anisotropic effects grows with the increase of the third spatial distance \(|z|\).
Thus, the statistical anisotropy would be significant at large scales.
At small scales, however, the anisotropic effects are not significant in this model.
Note that the above primordial power spectrum (\ref{approximate parametrization}) could be constrained by the Planck's results
via a similar analysis in the reference \cite{Ma1102,CMB statistics for a direction-dependent primordial power spectrum}.

As an example, we refer to the linear approximation of \(B(z)\) in (\ref{B(z) solution}), i.e., \(B(z)=f_{1}z+o(z^2)\).
Here \(f_{1}\) is a constant coefficient with the dimension \([length]^{-1}\).
Thus, \(B\) becomes
\begin{equation}
\label{B(z)f1}
B(z)=f_{1}\textbf{r}\cdot\hat{\textbf{n}}\ .
\end{equation}
By substituting the above equation into (\ref{approximate parametrization}),
one could obtain the primordial power spectrum as
\begin{equation}
\label{p1}
\mathcal{P}_{\mathcal{R}}(\textbf{k},\textbf{r})\equiv \mathcal{P}_{\mathcal{R}}^{iso}(k)(1-3f_{1}\hat{\textbf{n}}\cdot\textbf{r}
+6f_{1}^{2}(\hat{\textbf{n}}\cdot\textbf{r})^{2}-\frac{27f_{1}^{2}}{8k^{2}}(\hat{\textbf{n}}\cdot\hat{\textbf{k}})^{2})\ .
\end{equation}
The last two terms refer to the quadrupole asymmetry in the above equation.
Generically, they are smaller than the dipolar term.
Thus, one could ignore them first and only interested on the dipolar modulation of the power spectrum.
In this way, the primordial power spectrum (\ref{p1}) could be rewritten as
\begin{equation}
\label{dipolar modulation}
\mathcal{P}_{\mathcal{R}}(\textbf{k},\textbf{r})=\mathcal{P}_{\mathcal{R}}^{iso}(k)\left(1-3f_{1}\hat{\textbf{n}}\cdot\textbf{r}\right)\ .
\end{equation}
According to Eq.~(\ref{primordial power spectrum with direction dependence}),
this dipolar modulation of primordial power spectrum could account for the hemispherical asymmetry
of the CMB power observed by the WMAP and Planck satellites.

In the following, we estimate the amount, scale, and direction of the anisotropic effect based on the Planck results.
First, the primordial power spectrum (\ref{dipolar modulation}) could be parameterized as
\begin{equation}
\label{final primordial power spectrum}
\mathcal{P}_{\mathcal{R}}(\textbf{k},\textbf{r})=\mathcal{P}_{\mathcal{R}}^{iso}(k)\left(1+\frac{k_{c}}{k}\hat{\textbf{n}}\cdot\hat{\textbf{r}}\right)\ ,
\end{equation}
where we have used \(r\sim k^{-1}\), and several extra constants have been absorbed into the parameter \(k_{c}\).
The constant \(k_{c}\) denotes a critical wavenumber, which refers the critical scale of the dipolar modulation.
Thus, we obtain the magnitude of the anisotropic effect, i.e., \(A=k_{c}k^{-1}\).
Based on the Planck 2013 results \cite{Planck2013resultsXXIII}, the parameter \(A\) is
of level \(\sim0.07\) at the scale \(k^{-1}\sim 1\rm{Gpc}\).
We could get the parameter \(k_{c}^{-1}\sim 14\rm{Gpc}\), which is comparable to the scale of the universe.
This estimation implies that the anisotropic term would affect the CMB physics only at the low--\(\ell\) multipoles.
At small scales, the parameter \(A\) would decay rapidly with respect to \(k\).
This is consistent with the constraint on \(A\) from the distribution of distant quasars \cite{Constraints on cosmic hemispherical power anomalies from quasars}.
In addition, the anisotropic axis \(\hat{\textbf{n}}\) orients to the direction \((l,b)=(218.9^\circ \pm15.4^\circ,-21.4^\circ \pm15.1^\circ)\).

\section{Conclusions and discussions}\label{sec:Conclusions}

In this paper, we have studied a generalized FRW metric with weak anisotropy in the Randers spacetime.
The osculating Riemannian approach was employed to obtain the evolution of the Randers spacetime.
We found an inflationary solution of Einstein's gravitational field equations at zero order.
The anisotropic modifications were obtained to the inflationary stage of the universe.
Most importantly, the primordial power spectrum was obtained by analysis on the equation of motion for the inflaton perturbations.
It was found to acquire the dipolar modulation along the third spatial axis.
This could reveal the dipolar power modulation of the CMB temperature fluctuations.
The anisotropic effects are significant at large--angular scales while they are insignificant at small--angular scales in this model.
The primordial power spectrum with direction dependence could account for certain anomalies of the CMB temperature fluctuations
\cite{nonstandard spinor,curvaton field01,curvaton field02,DurrerKY1998,KimN2009,vector field02,vector field03,Cosmological Magnetic Fields01,
Cosmological Magnetic Fields02,Cosmological Magnetic Fields03,Cosmological Magnetic Fields04,Cosmological Magnetic Fields05,
primordial preferred direction001,preferreddirectioninWMAP5evidence01,preferreddirectioninWMAP5evidence02,preferreddirectioninWMAP5evidence03,
The statistically anisotropic curvature perturbation 1302,Ellipsoidal Universe01,Ellipsoidal Universe02,Large Scale Anisotropic Bias,
Spontaneous isotropy breaking,Direction dependence of the power spectrum1302}.

Based on the Planck 2013 results, we estimated the magnitude, scale, and direction of
the anisotropic effect in the Randersian inflation.
The Planck data showed the magnitude of the dipolar power modulation to be the level of \(A\sim0.07\) at the scale \(\sim1\rm{Gpc}\).
Thus, we obtained the scale of the anisotropic effect as \(k_{c}^{-1}\sim14\rm{Gpc}\).
This scale is comparable to the scale of the observable universe.
This prediction reveals that the anisotropic term affects the CMB anisotropy only at the low--\(\ell\) multipoles.
This is consistent with the WMAP and Planck observations on the CMB temperature fluctuations.
The anisotropic effect decays rapidly at small scales.
This is compatible with the constraint on the homogeneity from the distribution of distant quasars.
In addition, we regained the privileged axis of the dipolar modulation
\((l,b)=(218.9^\circ \pm15.4^\circ,-21.4^\circ \pm15.1^\circ)\) based on the Planck data.

As we have mentioned above, Finsler geometry is a reasonable candidate platform to study the spacetime asymmetry and anisotropy.
The Finsler spacetime is intrinsically anisotropic and direction dependent.
The physical processes in Finsler spacetime should acquire similar anisotropic properties.
In this paper, we indeed obtianed the anisotropic effects on the inflation and the primordial power spectrum from Finsler geometry.
To our knowledge, this is the first time that the primordial power spectrum was obtained in the Finsler geometric framework.
In addition, the statistical anisotropy was predicted for the primordial power spectrum, especially the dipolar modulation.
Meanwhile, the Planck's observations showed possible evidence for the dipolar power modulation of the CMB anisotropy at low--\(\ell\) multipoles.
This is a good chance to test the anisotropic predictions of Finsler geometry.
On the contrary, Finsler geometry could provide an alternative explanation to the statistical anisotropy of the universe.\\

\begin{acknowledgments}
We would like to thank Yunguo Jiang, Miao Li, Ming-Hua Li, Xin Li and Hai-Nan Lin for helpful discussions. Meanwhile, we are grateful to Danning Li for his help in our analytic calculations. Thanks David Lyth for his useful comment. This work has been funded by the National Natural Science Fund of China under Grant No. 11075166 and No. 11147176.
\end{acknowledgments}

\end{document}